\documentclass[letterpaper]{JHEP3}
\usepackage{epsfig,axodraw} 

\newcommand{\one}{{(1,0)} }
\newcommand{\MBH}{ M_B}

\newcommand{\be}{\begin{equation}}
\newcommand{\ee}{\end{equation}}
\newcommand{\bear}{\begin{eqnarray}}
\newcommand{\eear}{\end{eqnarray}} 
\newcommand{\ba}{\begin{array}}
\newcommand{\ea}{\end{array}}

\title{\Large Spinless photon dark matter from two universal extra dimensions 
{ $\; $ } \\ }

\author{Bogdan A.~Dobrescu$^1$, Dan Hooper$^2$, Kyoungchul Kong$^1$, Rakhi Mahbubani$^1$ 
\\ \\ 
$^1$Theoretical Physics Department, Fermilab, Batavia, IL 60510, USA \\ 
$^2$Theoretical Astrophysics Department, Fermilab, Batavia, IL 60510, USA \\ \rule{0em}{0.6em}
\email{bdob@fnal.gov, dhooper@fnal.gov, kckong@fnal.gov, rakhi@fnal.gov} \\ 
}

\abstract{We explore the properties of dark matter in theories with two universal
extra dimensions, where the lightest Kaluza-Klein state is a spin-0 neutral particle,
representing a six-dimensional photon polarized along the extra dimensions.
Annihilation of this `spinless photon' proceeds predominantly through Higgs boson 
exchange, and is largely independent of other Kaluza-Klein particles. 
The measured relic abundance sets an {\it upper} limit on the spinless photon mass
of 500 GeV, which decreases to almost 200 GeV if the Higgs boson is light.
The phenomenology of this dark matter candidate is strikingly different
from Kaluza-Klein dark matter in theories with one
universal extra dimension.
Elastic scattering of the spinless photon with quarks is helicity suppressed, 
making its direct detection challenging, although possible at upcoming experiments.
The prospects for indirect detection
with gamma rays and antimatter are similar to those of neutralinos.  
The rates predicted at neutrino telescopes are below the 
sensitivity of next-generation experiments.
  \\ \\ }

\preprint{{\small FERMILAB-PUB-07-274-A-T} \\ October 3, 2007  } 

\begin{document}

\section{Introduction} \setcounter{equation}{0}

Theories with universal extra dimensions~\cite{Appelquist:2000nn} have a $Z_2$ symmetry,
which is a remnant of invariance under translations along the compact dimensions.
This $Z_2$ symmetry, usually called Kaluza-Klein (KK) parity, implies that the lightest
KK particle is stable, and a potentially viable dark matter candidate. 

In the case of a single universal extra dimension compactified on an interval,
the geometrical origin of KK parity is the invariance under reflections with 
respect to the center of the interval. A one-loop computation of the mass splitting between 
KK particles shows that the lightest KK particle is typically the level-1 mode of the 
hypercharge gauge boson~\cite{Cheng:2002iz}. It turns out that this is  
an attractive dark matter candidate~\cite{Servant:2002aq, Cheng:2002ej,Hooper:2007qk}, 
whose relic abundance is consistent with the observed dark matter density for a mass between  
500 GeV and about 1.5 TeV, as shown by detailed computations including
coannihilations~\cite{Kong:2005hn} and 
level-2 resonances~\cite{Kakizaki:2005en}. 
Direct detection of this KK dark matter is possible with next 
generation experiments~\cite{Cheng:2002ej,Servant:2002hb,Majumdar:2002mw}, 
while indirect detection has somewhat better 
prospects than is found in the case of neutralinos~\cite{Cheng:2002ej,Hooper:2002gs,Bertone:2002ms,kribspos,Bergstrom:2004cy}.
Other dark matter candidates, such as the level-1 KK mode of the graviton or of a right-handed neutrino, 
are also viable for certain ranges of parameters in models with one universal 
extra dimension~\cite{Feng:2003xh,Cembranos:2006gt,Matsumoto:2006bf}.

Theories with two universal extra dimensions (see Ref.~\cite{Burdman:2006gy} and references therein)
also contain a KK parity. In the case of the simplest compactification 
that leads to chiral zero-mode fermions, a (`chiral') square with 
adjacent sides identified ~\cite{Burdman:2005sr,Dobrescu:2004zi},
the KK parity transformations are reflections with respect to the 
center of the square. Momentum along the two compact dimensions is 
quantized such that any 6-dimensional field propagating on the square
appears as a set of 4-dimensional particles labeled by two positive 
integers, $(j,k)$. These particles are odd under KK parity when
$j+k$ is odd, and are even otherwise. In any process, odd particles 
may be produced or annihilated only in pairs. 
The lightest odd particle, which is one of the (1,0) states, is thus stable.

Gauge bosons propagating in six dimensions may be polarized along the 
two extra dimensions. As a result, for each spin-1 KK particle
associated with a gauge boson, there are two spin-0 fields
transforming in the adjoint representation of the gauge group. 
One linear combination becomes the longitudinal degree of freedom of the
spin-1 KK particle, while the other linear combination remains as a
physical spin-0 particle, called the spinless adjoint. 

The 6-Dimensional Standard Model (6DSM), in which the Standard Model fields
and three right-handed neutrinos propagate in two universal extra dimensions
compactified on the chiral square, has been described in Ref.~\cite{Burdman:2006gy}.
Including one-loop corrections to masses in the 6DSM~\cite{Ponton:2005kx},
the lightest (1,0) particle is a linear combination of the electrically-neutral 
spinless adjoints of the electroweak gauge group. This is essentially a photon
polarized along the extra dimensions, which we will refer to as the 
`spinless photon'. At colliders, (1,0) particles may be pair produced and then undergo
cascade decays that end with spinless photons escaping the detector 
\cite{Dobrescu:2007xf}.

In this paper we study the viability of the spinless photon as dark matter,
as well as the prospects for its detection.
In the absence of majorana masses, the scalar nature of this dark matter candidate implies that its scattering cross sections with 
Standard Model fermions are suppressed, being proportional to the fermion mass. 
This is in contrast to the case of KK dark matter in one universal extra dimension,
where the lightest KK particle has spin 1, which allows for a large annihilation cross sections
to leptons. Nevertheless, the spinless photons may annihilate into $W^+W^-$, $ZZ$ and Higgs 
boson pairs, and we will show that for a range of masses correlated with the Higgs mass,
the relic abundance is 
consistent with the measured dark matter abundance. Although elastic scattering of spinless 
photons with nucleons is 
similarly helicity suppressed, its direct detection may be possible 
at next-generation experiments. The relatively small elastic 
scattering cross section leads to undetectable rates at neutrino telescopes.
Furthermore, given that 
pairs of spinless photons annihilate into heavy Standard Model particles, 
their indirect detection with gamma rays and antimatter is somewhat more difficult
than in the 5D case. We find that 
in most phenomenological respects, dark matter in the 6DSM more closely
resembles a neutralino than KK dark matter in one universal extra dimension.

\newpage
\section{Spinless photon annihilation} \setcounter{equation}{0}

The mass spectrum of (1,0) particles in the 6DSM \cite{Burdman:2006gy}, including the logarithmically 
enhanced one-loop corrections computed in Ref.~\cite{Ponton:2005kx}, is detailed in 
Ref.~\cite{Dobrescu:2007xf}. The essential feature of that spectrum is that the spinless adjoint
of the hypercharge gauge group, $B_H^\one$ (labeled for brevity $B_H$ in this paper),
is the lightest (1,0) particle, and therefore a dark matter candidate. 

There may be contributions from cutoff-scale physics
to operators localized at the corners of the square compactification, 
which are invariant under KK parity and modify the mass spectrum
\cite{Burdman:2006gy}. In principle, these could turn some other (1,0) 
particle into the lightest KK-odd state. Hence, the (1,0) modes of the 
graviton ($ {\cal G}_{\mu\nu}^\one $), 
of the right-handed neutrinos ($N_-^\one$), 
of one of the electrically-neutral components of the Higgs doublet ($H^{\one 0}$)
or of the electroweak bosons ($B_{\mu}^\one$, $W_{\mu}^{\one 3}$, $W_{H}^{\one 3}$), 
could all be viable dark matter candidates. 
We leave the investigation of these possibilities for future work.

\begin{table}[t]
\begin{center}
\renewcommand{\arraystretch}{1.7}
\begin{tabular}{|c|c||c|c|}
\hline \ \hspace*{-.28cm}boson\hspace*{-.28cm} \ & $M R$ \ 
& \hspace*{-.16cm}fermion\hspace*{-.17cm} & $M R$ \
\rule{0mm}{5mm}\rule{0mm}{-22mm} \\ \hline\hline 
$G_\mu^\one$ , \  $G_H^\one$  & 1.39 , \ 1.00  & $\left(T_+^{\one},B_+^{\one}\right)$ & 
$1.27 + \frac{1}{2}(m_t R)^2$  \\ \hline
\ $W_\mu^{\one 3}$ , \ $W_\mu^{\one \pm}$ \ & \ $1.06  + \frac{1}{2}(m_W R)^2$ & $T_-^\one$ \ & 
\ $1.25 +  \frac{1}{2}(m_t R)^2$ \ \\ \hline
$\left(H^{\one +},H^{\one 0}\right)$ & $1.05 + \Delta_h$ &  $\left(U_+^\one,D_+^\one\right)$ & 1.25 \\ \hline 
${\cal G}_{\mu\nu}^\one$, $B_\mu^\one$ & 1.00 , \ 0.97 & \ $U_-^\one$ , \ $D_-^\one$ \ & 1.22 , \ 1.21 \\ \hline
$W_H^{\one 3}$ , \ $W_H^{\one \pm}$  & $0.92 + \frac{1}{2}(m_W R)^2$  & $\left(N_+^\one,E_+^\one\right)$ & 1.04 \\ \hline
$B_H^\one \equiv B_H$   & 0.86 & $E_-^\one$ , \  $N_-^\one$ & 1.04 , \ 1.00  \\ \hline
\end{tabular} 
\vspace*{.3cm}
\caption{Masses of the (1,0) particles in units of the compactification scale $1/R$. 
The (1,0) fermion masses are almost the same for all three generations, with the exception of the 
top-quark KK modes.  The 
mass splittings depend on standard model couplings, and thus depend logarithmically 
on $1/R$. Here we used $1/R = 500$ GeV, and we kept only the leading terms
in the $m_t R$ expansion, where $m_t$ is the top-quark mass.
The correction $\Delta_h$ to the (1,0) Higgs masses is unknown, being quadratically sensitive to the cutoff scale.
}
\label{tab:mass} 
\end{center}
\end{table}

Electroweak symmetry breaking induces mixing between $B_H$ and the electrically-neutral 
spinless adjoint of $SU(2)_W$, $W_H^{\one 3}$, so that it is appropriate to call $B_H$ the spinless
photon. However, this mixing is 
suppressed by $m_W R$, where $m_W$ is the $W$ boson mass, and $1/R$ is 
the compactification scale.
For simplicity we will ignore mixing effects in what follows. This approximation is not valid if both
$M_B$ and the mass of $W_H^{\one 3}$ are below ${\cal O}(100)$ GeV. However, localized operators could
increase  the mass of $W_H^{\one 3}$ without changing $M_B$, so in the limit where 
$W_H^{\one 3}$ is much heavier than $m_W$ our results apply to any value of $M_B$.

As we will see in this section, the only 
other (1,0) particles that affect the annihilation cross section of $B_H$ are the KK modes of the 
top quark: $T_-^\one$, which is an $SU(2)_W$-singlet vectorlike quark, 
and $T_+^\one$, which together with $B_+^\one$ forms an 
$SU(2)_W$-doublet vectorlike quark. The masses of other (1,0) quarks are necessary 
for computing  the elastic scattering cross section of $B_H$ with nucleons (see Section 4). 
The masses of the (1,0) leptons and vector bosons are largely irrelevant for our present study.
Nevertheless, we show in Table~\ref{tab:mass} the full (1,0) spectrum from Ref.~\cite{Dobrescu:2007xf}, 
which turns out to include sufficiently large mass splittings so that coannihilation effects 
may be neglected.  We loosely refer to all (1,0) particles as `level-1' modes in what follows, and 
we label them using the superscript $\one$.

\subsection{Annihilation into boson pairs}

The interaction of the $B_H$ with the Standard Model Higgs boson, $h$, is given by
\begin{equation}
{\cal L}_{h} = - \frac{g_Y^2}{8} B_H B_H h \left( h + 2 v\right) \, ,
\end{equation}
where $g_Y$ is the hypercharge gauge coupling and $v \approx 246$ GeV 
is the electroweak scale.
There are no tree-level interactions of the type  $B_H H^\one h$, 
$\partial_\mu B_H H^{\one 0} Z^\mu$, or $\partial_\mu B_H H^{\one \mp} W^{\mu \pm}$.

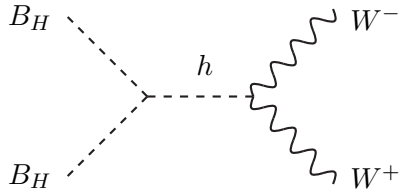
\begin{figure*}[t]
\unitlength=1.0 pt
\SetScale{1.0}
\SetWidth{0.8}      
\begin{center}
\begin{picture}(150,80)(0,20)
\DashLine(50,50)( 20,20){3}
\DashLine(  20,80)( 50,50){3}
\DashLine(50,50)(90,50){3}
\Photon(120,17)(90,50){-3.5}{4}
\Photon(120,83)(90,50){3.5}{4}

\Text(146,20)[r]{$W^+ $}
\Text(146,80)[r]{$W^- $}
\Text(14,20)[r]{$B_H$}
\Text(14,80)[r]{$B_H$}
\Text(75,62)[r]{$h$}
\end{picture}
\end{center}
%
\caption{The only tree-level contribution to $B_H B_H $ annihilation into $W^+W^-$.
The same diagram with the $W$ bosons replaced by $Z$ bosons 
describes annihilation into $Z$ pairs.}
\label{fig:WW}
\end{figure*}
The annihilation cross section into a $W^+W^-$ pair (see Fig.~\ref{fig:WW}) is given by
\be
\sigma (B_H B_H \to W^+ W^-) =
\frac{ g_Y^4 ( s^2 - 4 m_W^2 s +  12  m_W^4 ) }
     { 64 \pi s \left(  s - m_h^2 \right)^2 } \left(\frac{s - 4m_W^2}{s - 4\MBH^2}\right)^{\! 1/2} ~,
\ee
and the same expression with the $W$ boson mass replaced by the $Z$ boson mass yields the cross section for $B_H B_H$ annihilation into a $ZZ$ pair
\be
\sigma (B_H B_H \to Z Z) =
\left. \frac{1}{2} \; \sigma (B_H B_H \to W^+ W^-)\right|_{m_W \rightarrow m_Z}   ~,
\ee  
where the factor of 1/2 results from having two identical particles in the final state. 
Here $s$ is the center-of-mass energy of the collision, while $m_W$, $m_Z$ and $m_h$ are the 
the Standard Model masses.

Expanding the cross section in powers of the relative speed between the $B_H$ bosons, $v_r$, gives 
\be
v_r \, \sigma\!\left( B_H B_H \rightarrow  W^+ W^- \right) 
= a_W + v_r^2 b_W +{\cal O}\left(v_r^4\right) ~.
\ee
The first two terms in this non-relativistic expansion are
\be
a_{W} = \frac{2\pi \alpha^2  \MBH^2 }{c_w^4 \left(4 \MBH^2-m_h^2\right)^2}
\left(1 -  \frac{m_W^2}{\MBH^2} + \frac{3 m_W^4}{4\MBH^4} \right) 
\left(1 - \frac{m_W^2}{\MBH^2}\right)^{\! 1/2} ~~,
\label{aW}
\ee
and 
\be
b_{W} = \frac{- a_W}{4\left(\MBH^2 - m_W^2\right)}
\left( \MBH^2\frac{4 \MBH^2 + 3 m_h^2 - 16 m_W^2}{2\left(4 \MBH^2 - m_h^2\right)}
+  \frac{3 m_W^4\left(2 \MBH^2 - m_W^2\right)}{4 \MBH^4 - 4 \MBH^2 m_W^2 + 3 m_W^4}
\right) ~,
\ee
where $\alpha$ is the fine structure constant evaluated at the scale 
$M_B$ and $c_w = \cos\theta_w$ is the cosine of the weak mixing angle.

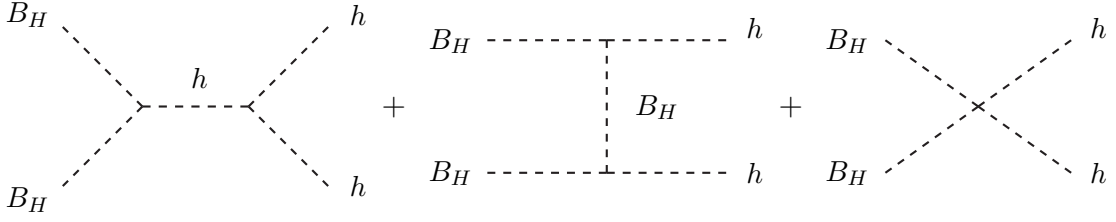
\begin{figure*}[t]
\unitlength=1.0 pt
\SetScale{1.0}
\SetWidth{0.8}      
\begin{center}
\begin{picture}(405,80)(0,20)
\DashLine(50,50)(20,20){3}
\DashLine(20,80)(50,50){3}
\DashLine(50,50)(90,50){3}
\DashLine(120,20)(90,50){3}
\DashLine(90,50)( 120,80){3}
\Text(150,50)[r]{{\large +}}
\DashLine(180,75)(225,75){3}
\DashLine(225,75)(225,25){3}
\DashLine(225,25)(180,25){3}
\DashLine(270,75)(225,75){3}
\DashLine(225,25)(270,25){3}
\Text(300,50)[r]{{\large +}}
\DashLine(330,75)(365,50){3}
\DashLine(330,25)(365,50){3}
\DashLine(400,75)(365,50){3}
\DashLine(400,25)(365,50){3}
\Text(15,15)[r]{$B_H$}
\Text(15,85)[r]{$B_H$}
\Text(175,25)[r]{$B_H$}
\Text(175,75)[r]{$B_H$}
\Text(325,25)[r]{$B_H$}
\Text(325,75)[r]{$B_H$}
\Text(75,61)[r]{$h$}
\Text(135,20)[r]{$h$}
\Text(135,85)[r]{$h$}
\Text(285,25)[r]{$h$}
\Text(285,80)[r]{$h$}
\Text(415,25)[r]{$h$}
\Text(415,80)[r]{$h$}
\Text(253,50)[r]{$B_H$}
\end{picture}
\end{center}
%
\caption{Tree level diagrams for $B_H B_H$ annihilation into $hh$ (the $u$-channel diagram is not shown).}
\label{fig:hh}
\end{figure*}
The annihilation cross section into a $hh$ pair (see Fig.~\ref{fig:hh}) is given by 
\bear
\sigma (B_H B_H \to h h) &=&
\frac{g_Y^4}{16\pi s} \left[ \left( \frac{(s+2 m_h^2)^2}{8 (s-m_h^2)^2}  + 
       \frac{m_Z^4 s_w^4}{m_h^4+\MBH^2 (s-4m_h^2) } \right)
      \left ( \frac{s-4m_h^2}{s-4 \MBH^2} \right ) ^{1/2} \right.
\nonumber \\ [0.7em]
&&\hspace*{-3cm} + \left. \,\frac{m_Z^2 s_w^2}{s-4\MBH^2} 
\left ( \frac{s+2m_h^2}{s-m_h^2} - \frac{2m_Z^2 s_w^2}{s-2m_h^2}
\right ) \ln \left ( \frac{s- 2m_h^2- \sqrt{(s-4\MBH^2)(s-4m_h^2) }}
                           {s- 2m_h^2+ \sqrt{(s-4\MBH^2)(s-4m_h^2) }}\right ) \right]~. 
\eear 
The corresponding leading terms in the non-relativistic expansion are
\be
a_{h} = \frac{\pi \alpha^2 \sqrt{\MBH^2 - m_h^2}}{4 c_w^4 \MBH^3 }
\left ( \frac{2 \MBH^2 + m_h^2}{4 \MBH^2 - m_h^2} + \frac{2 m_Z^2 s_W^2}{2 \MBH^2 - m_h^2}  \right )^2
\ee
and
\begin{eqnarray}
b_{h} & = & \frac{a_h}{ 2\MBH^2 + m_h^2} \left( - \frac{8\MBH^6 + 10 \MBH^4 m_h^2 - 29 \MBH^2 m_h^4 + 2m_h^6}
{8 \left(4 \MBH^2 - m_h^2\right)\left(\MBH^2 - m_h^2\right)} \right.
\nonumber \\ [0.7em]
&& +\; \left. \frac{4}{3} \MBH^2 M_Z^2 s_w^2 \frac{16\MBH^6 -18 \MBH^4 m_h^2 + 15 \MBH^2 m_h^4 -4 m_h^6}
{\left(2\MBH^2 - m_h^2\right)^2\left[4 \MBH^4 - m_h^4 - 2 M_Z^2 s_w^2 \left(4 \MBH^2 - m_h^2 \right) \right]} \right) ~~.
\label{bh}
\end{eqnarray}
In the limit in which all the Standard Model particles are much lighter than $B_H$, 
the equivalence theorem holds for the boson final states:
\begin{equation}
\sigma_{hh} = \sigma_{ZZ} = \frac{1}{2} \sigma_{W^+W^-} =  \frac{g_Y^4}{256 \pi M_B^2 v_r} 
\left( 1 - \frac{1}{8} v_r^2 + \cdots \right) \, .
\end{equation}
%
%
\subsection{Annihilation into fermion pairs}\label{sec:xsec}

On general grounds, the interaction between 
a pair of $B_H$ particles and a pair of fermions is helicity suppressed.
To see this note that operators that include a derivative, such as
\bear
{\cal O}_1 & = & \frac{i}{\Lambda^2} B_H B_H \bar{f} \, \gamma^\mu \partial_\mu f ~,
\nonumber \\ [0.4em]
{\cal O}_2 & = & \frac{1}{\Lambda^2} B_H \left(\partial_\mu B_H\right) \, \bar{f} \gamma^\mu \gamma_5 f ~~,
\eear
may be integrated by parts, and then using the Dirac equation take the equivalent form
\bear
\label{equ:ops}
{\cal O}_1 & = & \frac{m_f }{\Lambda^2} B_H B_H \, \bar{f} f ~~,
\nonumber \\ [0.4em]
{\cal O}_2 & = & -\frac{i m_f}{\Lambda^2} B_H  B_H \, \bar{f}\gamma_5 f \, ~.
\eear
Thus, the two above operators, suppressed by the ratio of the fermion mass $m_f$ to some cutoff scale $\Lambda$,
are the only independent Lorentz-invariant operators 
that describe the interactions of two $B_H$'s with a fermion-antifermion pair.
These operators are written in an effective theory below the electroweak scale. However, the same arguments
apply when the operators are written in an $SU(2)_W\times U(1)_Y$-invariant way, with $m_f$ replaced by $\lambda_f H$
where $\lambda_f$ is the Yukawa coupling of the fermion to the standard model Higgs doublet $H$.

The two operators shown in Eq.~(\ref{equ:ops})
govern the annihilation of spinless photon dark matter to fermions as well as its elastic 
scattering with nucleons. Hence both these processes
will be suppressed by standard model fermion masses.  
In the 6DSM there are contributions to the operators in Eq.~\ref{equ:ops} 
from Higgs exchange and (1,0) quark exchange.  Higgs exchange 
contributes only to ${\cal O}_1$, whereas KK quark exchange can contribute to both operators.  
Therefore, the cutoff scale $\Lambda$ is given in practice by either the mass of a KK quark 
or by the Higgs boson mass. 
We will verify these statements by explicit computation
of cross sections below, focusing on annihilation to top quarks.

The interaction between the $B_H$ and top quarks takes the following 
form:
\begin{eqnarray}
{\cal L}_{t} &=&  i \frac{g_Y}{2}  B_H\left( y_L \, \bar{T}_{+_R}^\one t_L 
                       + y_R  \, \bar{T}_{-_L}^\one t_R \right) + {\rm H.c.}        \, ,
\end{eqnarray}
where $y_L = 1/3$ and $y_R = 4/3$ are the hypercharges of left-handed and right-handed top quark and 
$P_{L/R} = (1 \mp\gamma_5)/2$ is the projection operator. 
Interactions with the Standard Model Higgs boson generates 
off-diagonal elements in
the mass matrix of the level-1 top quarks after electroweak
symmetry breaking,
\begin{equation}
\left ( 
\begin{array}{cc}
\bar{T}_-^\one  & \bar{T}_+^\one 
\end{array}
\right )
\left ( 
\begin{array}{cc}
- \frac{1}{R} \left ( 1+ \Delta_- \right )  & m_t (1 + \delta_1) \\
 m_t (1 + \delta_2)                         &  \frac{1}{R} \left ( 1+ \Delta_+ \right )
\end{array}
\right )
\left ( 
\begin{array}{c}
{T}_-^\one  \\
{T}_+^\one 
\end{array}
\right )          \, ,
\end{equation}
where the $\delta$s and $\Delta$s are 
radiative corrections to the heavy quark masses.
The dominant contribution to these comes from the strong interaction and in the
limit that we ignore electroweak corrections,
$\Delta_+ = \Delta_-=\Delta$  and $\delta_1 = \delta_2=\delta$.
The diagonal correction $\Delta$ was computed in Ref.~\cite{Ponton:2005kx} to be equal to
\begin{equation}
\Delta = \frac{16}{3} \frac{g_s^2 }{8\pi^2}\log \left ( \Lambda R \right ) + \frac{m_t^2 R^2}{2} + {\cal O}\left ( \frac{g^2}{g_s^2}, \frac{g_Y^2}{g_s^2}, \frac{\lambda_t^2}{g_s^2} \right )\;,
\end{equation}
where $g_Y$, $g$ and $g_s$ are the $SU(3)_c \times SU(2)_W \times U(1)_Y$ gauge couplings, $\lambda_t$ is the top 
Yukawa coupling, $m_t$ is the Standard Model top quark mass, and $\Lambda$ is the cut-off scale. We take $\Lambda \approx 10/R$ based 
on naive dimensional analysis~\cite{Burdman:2006gy}. Although $\delta$ has not been computed, it is expected to be of the same order as 
$\Delta$, and we will take these to be equal for the remainder of this paper.

The weak eigenstates are related to mass eigenstates by 
\begin{equation}
\left ( 
\begin{array}{c}
T_-^\one \\
T_+^\one
\end{array}
\right ) =
\left ( 
\begin{array}{cc}
-\gamma_5 c_\alpha & s_\alpha \\
\gamma_5 s_\alpha  & c_\alpha 
\end{array}
\right )
\left (
\begin{array}{c}
{T'}_-^{\one} \\
{T'}_+^{\one}
\end{array}
\right )           \, ,
\end{equation}
where $c_\alpha = \cos\alpha$, $s_\alpha=\sin\alpha$ for a mixing angle $\alpha$ given by $\tan 2 \alpha = m_t R$.  
The mass eigenstates, ${T'}_-^{\one}$ and ${T'}_+^{\one}$ have the same
mass
\begin{equation}
M_T = \sqrt{ \frac{1}{R^2}  + m_t^2  } ( 1+ \Delta )    \, .
\end{equation}
In the mass eigenstate basis the $B_H$-top quark interaction
can be written as
\begin{equation}
{\cal L}_{t} = 
 i \frac{g_Y}{2}  B_H \left [ \bar{T'}_-^\one \left ( y_L P_L s_\alpha + y_R P_R c_\alpha \right ) t
                     + \bar{T'}_+^\one  \left ( y_L P_L c_\alpha + y_R P_R s_\alpha \right ) t
                     \right ] + {\rm H.c.} \nonumber     \, .
\end{equation}
Since we will only deal with the quark mass eigenstates, we will omit all 
primes in what follows.

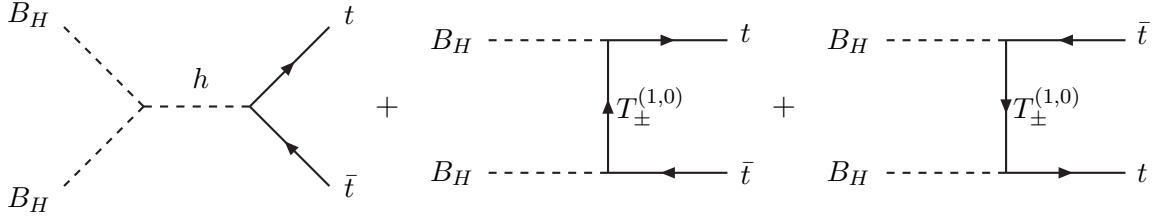
\begin{figure*}[t]
\unitlength=1.0 pt
\SetScale{1.0}
\SetWidth{0.8}      
\begin{center}
\begin{picture}(400,80)(10,20)
\DashLine(50,50)( 20,20){3}
\DashLine(  20,80)( 50,50){3}
\DashLine(50,50)(90,50){3}
\ArrowLine( 120,20)(90,50)
\ArrowLine(90,50)( 120,80)
\Text(147,50)[r]{{\large +}}
\DashLine(180,75)(225,75){3}
\DashLine(225,25)(180,25){3}
\ArrowLine(225,25)(225,75)
\ArrowLine(225,75)(270,75)
\ArrowLine(270,25)(225,25)
\Text(297,50)[r]{{\large +}}
\DashLine(330,75)(375,75){3}
\DashLine(375,25)(330,25){3}

\ArrowLine(375,75)(375,25)
\ArrowLine(375,25)(420,25)
\ArrowLine(420,75)(375,75)

\Text(15,15)[r]{$B_H$}
\Text(15,85)[r]{$B_H$}
\Text(175,25)[r]{$B_H$}
\Text(175,75)[r]{$B_H$}
\Text(325,25)[r]{$B_H$}
\Text(325,75)[r]{$B_H$}
\Text(75,61)[r]{$h$}

\Text(130,20)[r]{$\bar{t}$}
\Text(130,85)[r]{${t}$}
\Text(280,25)[r]{$\bar{t}$}
\Text(280,78)[r]{${t}$}
\Text(430,25)[r]{${t}$}
\Text(430,78)[r]{$\bar{t}$}

\Text(255,50)[r]{$T_\pm^\one$}
\Text(405,50)[r]{$T_\pm^\one$}
\end{picture}
\end{center}
%
\caption{Tree-level diagrams for $B_H B_H$ annihilation into $t\bar{t}$.}
\label{fig:TT}
\end{figure*}

The relativistic annihilation cross section into top quarks, computed at tree level,
is given by
\be
\sigma\left(B_H B_H \rightarrow t\bar{t} \right) =  
\frac{3 g_Y^4}{8 \pi s \left(s - 4 \MBH^2 \right)} 
\int_{t_-}^{t_+} \!\! dt \;
\left[ - \frac{y_L^4 + y_R^4}{16} B(t) +
\frac{y_L y_R}{8} m_t^2 A_T(t) +  m_t^2 A_h(t) 
\right] \,,
\label{sigma-t}
\ee
where the terms collected in $B(t)$ are due to exchange of 
the gauge eigenstates of the $T$ quark, without interference terms,
\be
B(t) = \frac{t^2 + t \left( s - 2 \MBH^2 \right) + \left( M_H^2 - m_t^2\right)^2}
{\left(t-M_T^2\right)^2}  +
\frac{t^2 + t \left( s - 2 \MBH^2 - 2 m_t^2 \right) + \MBH^4 - m_t^4}
{\left(M_T^2 - \MBH^2 - m_t^2 + s/2\right)\left(t-M_T^2\right)} ~.
\ee
The contributions from an electroweak mass insertion on the $T$ quark line
are included in
\bear
A_T(t) &=&  \frac{-1}{\left(t-M_T^2\right)^2}\left\{
y_L y_R \left[ 2t - \left( 1 + \delta \right)^2 \left( s - 4 m_t^2 \right) \right]
+ 2\left(y_L^2 + y_R^2\right)\left( 1 + \delta \right) \left( t - \MBH^2 + m_t^2\right)\right\}
\nonumber \\ [0.7em]
&& - \;
\frac{y_L y_R \left[2\left(\MBH^2 - m_t^2\right) + \left( 1 + \delta \right)^2 \left( s - 4 m_t^2 \right)\right] + \left(y_L^2 + y_R^2\right)\left( 1 + \delta \right)\left( s - 4 m_t^2 \right) }
{\left(M_T^2 - \MBH^2 - m_t^2 + s/2\right)\left(t-M_T^2\right)}  ~.
\eear
Finally, $A_h(t)$ includes the contributions due to Higgs boson exchange, 
\be
A_h(t) = \frac{ 
-\left(y_L^2 + y_R^2\right)\left(t - \MBH^2 + m_t^2\right)
+y_L y_R \left( 1 + \delta \right) \left(s - 4 m_t^2\right)}
{2\left(s - m_h^2\right)\left(t - M_T^2 \right)}
+ \frac{s- 4 m_t^2}{4\left(s - m_h^2\right)^2} ~~.
\ee
The integration limits of the Mandelstam variable, $t$, are given by
\be
t_\mp = M_B^2 + m_t^2 - \frac{s}{2} \mp 
\frac{1}{2} \sqrt{\left( s - 4 \MBH^2 \right)\left( s - 4 m_t^2 \right)} \, .
\ee
After integrating 
over $t$ in Eq.~(\ref{sigma-t}), we find the following leading 
terms in the non-relativistic expansion for $\sigma v_r$
\bear
a_{t} &=& \frac{3 \pi \alpha^2}{4 c_w^4}\frac{m_t^2}{\MBH^3} \left(\MBH^2 - m_t^2\right)^{\!3/2}
\left( \frac{(y_L+y_R)^2 + 2y_Ly_R\delta}{M_T^2 + \MBH^2 - m_t^2} - \frac{2}{4\MBH^2 - m_h^2} \right)^2
~,
\label{at}
\eear
and 
\bear
b_{t} & = & - \frac{a_t}{24} \left[6
- \frac{\MBH^2}{\MBH^2 - m_t^2} 
\left(1 - \frac{4M_T^2}{M_T^2 + \MBH^2 - m_t^2} \right)^{\! 2} \right.
+  \frac{8 \MBH^2}{4\MBH^2 - m_h^2}  \\ [0.5em]
& \times  &  \left.
\frac{\left(y_L^2+y_R^2\right)\left(4\MBH^2 - m_h^2\right)^2  
+ 2\left(3M_T^2 + \MBH^2 - m_t^2\right)\left(4\MBH^2 - m_h^2\right) 
-12\left(M_T^2 + \MBH^2 - m_t^2\right)^2 }
{\left[(y_L+y_R)^2 + 2y_Ly_R\delta\right]\left(M_T^2 + \MBH^2 - m_t^2\right)\left(4\MBH^2 - m_h^2\right)  - 2 \left(M_T^2 + \MBH^2 - m_t^2\right)^2 }
  \right] ~\!\!\!. \nonumber
\label{bt}
\eear
This computation confirms that annihilation into $t\bar{t}$ is 
suppressed by $m^2_t/\MBH^2$ due to helicity flipping.  Note the relative minus sign in Eq.~\ref{at}
between the Higgs-exchange and heavy top exchange contibutions. This interference leads to further
suppression of this annihilation channel.

\bigskip
\section{Relic abundance} \setcounter{equation}{0}

We begin this section with a review of the standard calculation for the 
thermal relic abundance of a stable, massive particle~\cite{Srednicki:1988ce}. We then compute the relic
abundance for the spinless photon in order to determine the range of
$M_B$, the spinless photon mass, consistent with the observed
abundance of dark matter.

\subsection{From annihilation cross sections to relic abundance}
\label{sec:rel}

The relic abundance of $B_H$ is given by solving Boltzmann's equation 
for the evolution of its number density, $n$,
\begin{equation}
\frac{d n}{ d t} = -3 Hn - \langle \sigma v_r \rangle \left( n^2 - n^2_{\rm eq}\right) ~,
\end{equation}
where $H$ is the Hubble parameter,
$\langle \sigma v_r \rangle$ is the thermal average of the total annihilation cross 
section of $B_H$ times the relative velocity of the annihilating particles, 
and $n_{\rm eq}$ is their equilibrium number density. 

An approximate analytical solution can be found for early and late times. 
At temperatures 
substantially above the spinless photon mass ($T \gg M_B$) there are roughly as 
many $B_H$ particles as photons and $n_{\rm eq} \sim T^3$.
For temperatures below $M_B$ the equilibrium density is 
Boltzmann-suppressed and is given in the non-relativistic approximation by 
\begin{equation}
n_{\rm eq} = \left ( \frac{M_B\, T}{ 2 \pi} \right )^{\! 3/2} e^{-M_B/T} ~~.
\label{neq}
\end{equation}
As the temperature decreases still further the $B_H$ annihilation rate eventually
drops below the Hubble expansion rate so $B_H$ cannot remain in equilibrium and becomes
a thermal relic. From this point on, the total number of $B_H$ particles stays constant, with a number density diluted by the expansion of the universe.  The temperature at which this
takes place is known as the freeze-out temperature, $T_F$, and is roughly determined by 
equating the dark matter annihilation rate to the expansion rate of the universe
\begin{equation}
\langle \sigma v_r \rangle n\big|_{T=T_F} \sim H \, ,
\end{equation}
giving the following equation which can be solved iteratively for $T_F$:
\begin{equation}
\frac{M_B}{T_F} = \ln\left[c(c+2) \frac{3}{4\pi^3}\left(\frac{5 M_B\, T_F}{2 g_\ast}\right)^{\! 1/2} 
M_{\rm Pl} \langle \sigma v_r\rangle\bigg|_{T=T_F} \right]\ .
\end{equation}
Here, $M_{\rm Pl} = 1.22\times 10^{19}$ GeV is the Planck scale,
$g_\ast$ is the total number of effectively massless degrees of freedom 
at  the freeze-out temperature 
and 
$c$ is an $O(1)$ constant that is determined by comparing to numerical
solutions of the Boltzmann equation.  Note that because of its logarithmic dependence 
on mass and cross section, the ratio of the freeze-out temperature to the dark matter mass is 
relatively insensitive to these quantities.

In the non-relativistic limit the thermally averaged annihilation cross section can 
be expressed as
\be
\langle \sigma v_r \rangle = a +  6 \, b \, \frac{T_F}{M_B}  + \cdots~~,
\ee
where the $a$- and $b$-terms are sums 
over the contributions for $W^+W^-$, $ZZ$, $hh$ and  $t\bar{t}$ final states given in 
Eqs.~(\ref{aW})-(\ref{bh}), (\ref{at}) and (\ref{bt}).
Using this approximation one can match the early and late-time solutions to the 
Boltzmann equation to find the current $B_H$ density,
\begin{equation}
\Omega_{B_H} h^2 \approx \frac{1.04 \times 10^9 \; {\rm GeV}^{-1}}{M_{\rm Pl} \sqrt{g_\ast}}
 \frac{M_B/T_F}{a  + 3 \, b \, T_F/M_B }\ ,
\end{equation}
%
where the dimensionful constant in the numerator comes from factors of the current 
critical density and entropy density.  
A more careful treatment of this method~\cite{Srednicki:1988ce} 
results in additional sub-leading terms
which can be accounted for by the replacement $b\to b-a/4$ in the above formulas.

Note that the non-relativistic expansion fails near 
$s$-channel resonances and final state thresholds~\cite{Griest:1990kh} and the relic 
abundance in the vicinity of these must be calculated by alternative methods.  A 
treatment of resonances in models with one universal extra dimension can be found 
in Ref.~\cite{Kakizaki:2005en}.

\subsection{Prediction for the spinless photon mass}

For the remainder of this analysis we will ignore $\Delta$, the 
one-loop QCD correction to quark masses, since this quantity has
a negligible effect on our results.  Furthermore, since we have
no robust information on the exact value of the Higgs mass, we
take this to be a free parameter.  For Higgs masses near 
$2 M_B$, there is a resonance
effect from an $s$-channel Higgs going on shell.
Away from this resonance and all mass thresholds, 
the non-relativistic expansion of the annihilation cross section
is a valid approximation and the relic abundance can be computed 
analytically using the expressions for the 
annihilation cross sections given in Sec.~\ref{sec:xsec}.  These 
were verified using our implementation~\cite{web} of the 6DSM in {\tt CalcHEP}~\cite{Pukhov:1999gg}.

\begin{figure}[t]
\centerline{
\includegraphics[width=.5 \textwidth]{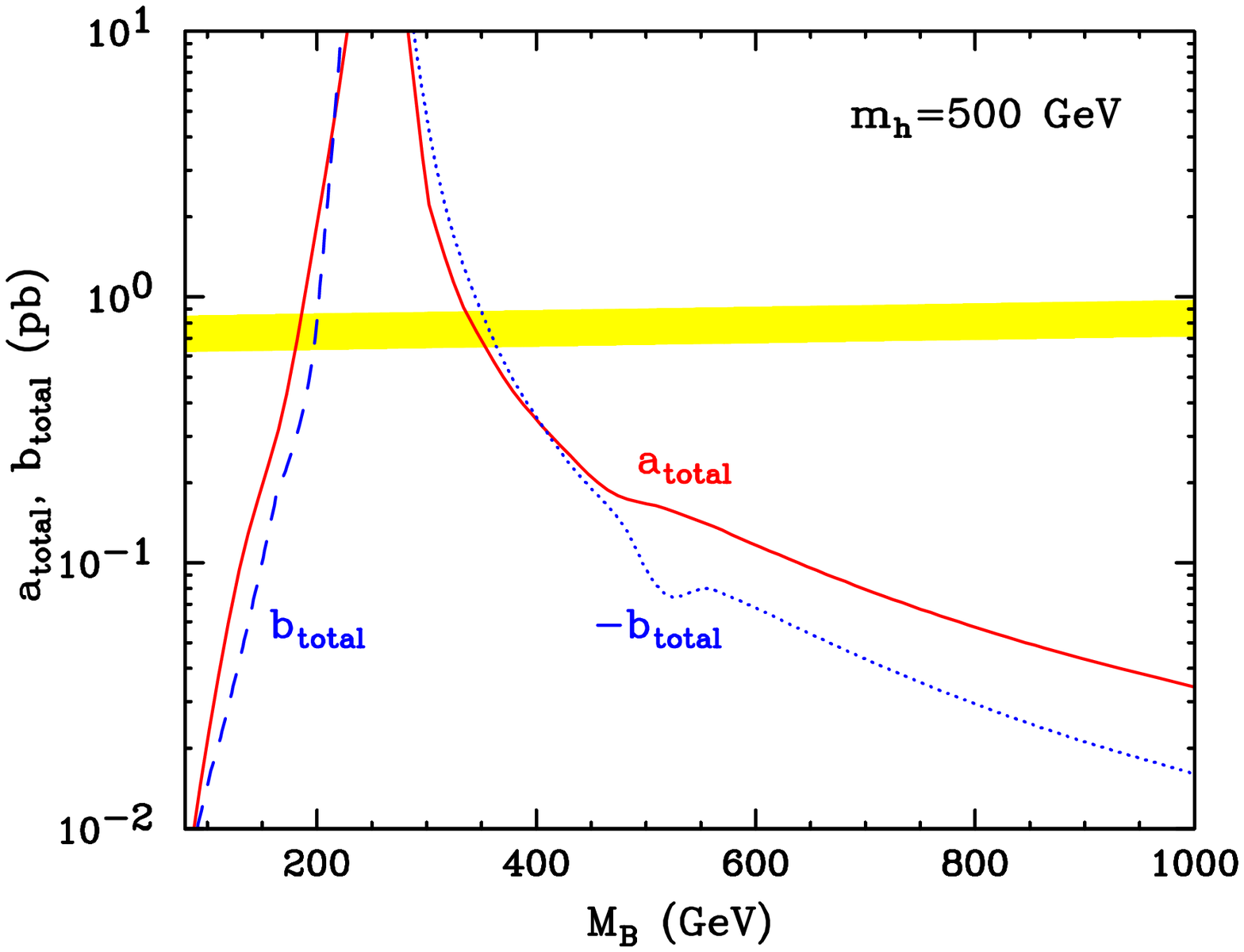} 
\includegraphics[width=.48 \textwidth]{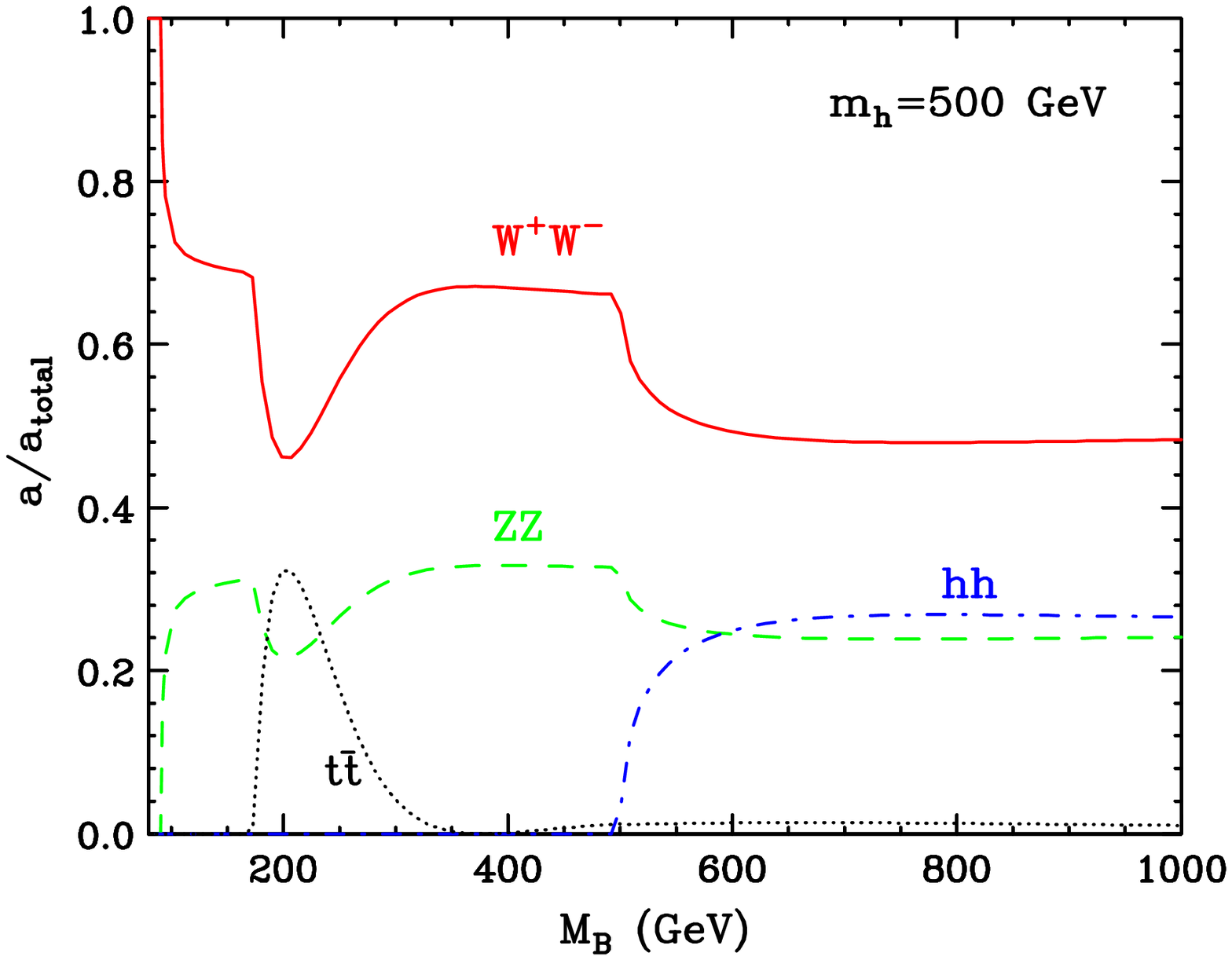} 
}
\caption{Left: The coefficients $a$ and $b$ from the non-relativistic expansion of 
the total $B_H$ annihilation cross section.  The shaded band corresponds to the current range
of $a$ measured by WMAP ($0.096 < \Omega_{B_H} h^2 < 0.122$ at 2$\sigma$) 
and $b$ includes the relativistic correction, $-a/4$.
Right: the relative contribution to $a_{\rm total}$ from various final states.
Note that the non-relativistic expansion fails near the Higgs $s$-channel resonance, 
$2 M_B=m_h=500$ GeV.}
\label{fig:xsection}
\end{figure}

In the left frame of Fig.~\ref{fig:xsection} we plot the $a$- and $b$-term contributions 
to the total annihilation cross section for a heavy Higgs boson, with the shaded
region corresponding to the range consistent with current WMAP data 
($0.096 < \Omega_{B_H} h^2 < 0.122$ at 2$\sigma$)~\cite{Spergel:2006hy}. 
In the regions away from the Higgs resonance, 
the total $b$-term is smaller than the 
$a$-term, although it becomes significant near the resonance due to the higher power of the mass
difference $4 M_B^2 - m_h^2$ in its denominator, in comparison with the $a$-term.
Even near the resonance, however, the effect of the $b$-term contribution on the relic abundance is suppressed by the velocity ($v_r^2  \sim 0.1$) and impacts the dark matter density at about the 10\% level or less. 

As shown in the left frame of Fig.~\ref{fig:xsection}, there are two regions consistent with WMAP 
around the Higgs resonance, $M_B\sim$ 180 GeV and $M_B\sim$ 350 GeV.
Note that in contrast to the 5D case~\cite{Servant:2002aq,Kong:2005hn} 
a light range of dark matter masses is preferred by data.  This difference is to a large 
extent due to the spin of the dark matter candidate. The dominant annihilation 
channel of the spin-1 dark matter candidate in 5D is to fermion pairs, whereas
annihilation of spinless photons to pairs of light fermions is helicity suppressed. 
The multiplicity of light fermion final states allows the former to annihilate more 
efficiently, leading to an increase in its mass in order to remain consistent with data.

\begin{figure}[t]
\centerline{
\includegraphics[width=.7 \textwidth]{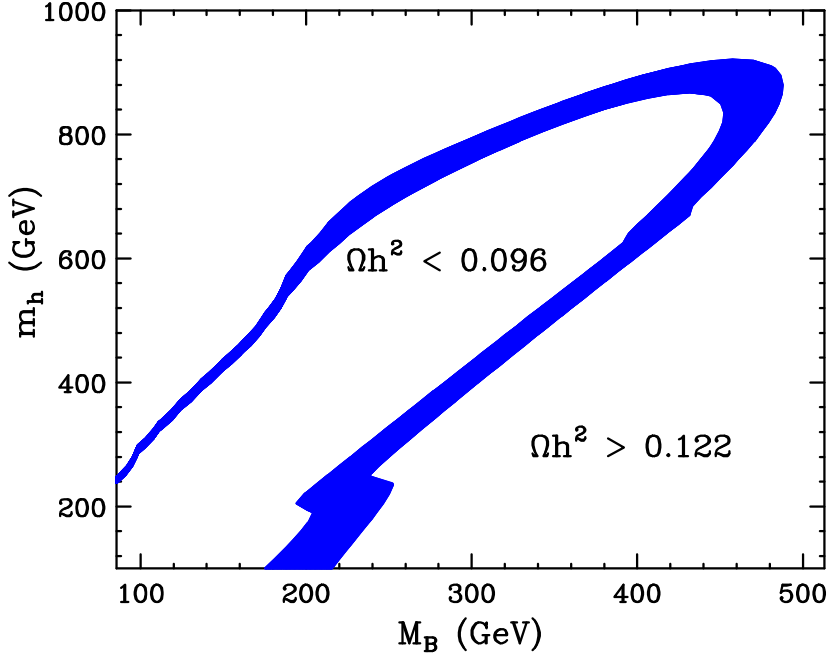} }
\caption{The region (shaded) of the $m_h$ vs. $M_B$ plane in which the $B_H$ thermal 
relic abundance is within the range measured by WMAP ($0.096 < \Omega_{B_H} h^2 < 0.122$).}
\label{fig:omegah2}
\end{figure}
The relative contributions to the total annihilation cross section 
from different final states are plotted for a large Higgs mass 
in the right frame of Fig.~\ref{fig:xsection}.
We see that annihilation to boson final states is dominant for a 
spinless 
photon mass above the boson production threshold. As expected from the 
Goldstone boson equivalence theorem, 
the $a$-term for the $W^+W^-$ final state is twice that for the $ZZ$ and $hh$ 
final states in the limit of large $M_B$. The top quark final state 
is only significant for a small range of parameters; it is 
below threshold for $M_B \lesssim$ 170 GeV and helicity suppressed for large 
values of $M_B$. 

Note that the results in this figure are not reliable in the region of $M_B 
\approx 250$ GeV as this 
corresponds to a spinless photon mass that is exactly half the Higgs mass
and the Higgs is on resonance. In such a case we can no longer use the
non-relativistic expansion of the annihilation cross-section, and instead calculate the relic abundance numerically using {\tt micrOMEGAs}
~\cite{Belanger:2006is}. Our results are shown in Fig.~\ref{fig:omegah2} for different
values of $m_h$ and $M_B$, with the shaded region corresponding to 
parameters that are consistent with the current WMAP measurements. 
In this figure we see again two possible regions of $M_B$ for each value of the Higgs mass, with the
region at smaller $M_B$ containing a significant contribution 
from annihilation to top pairs, this final state being helicity suppressed in the other region.
For instance, for a Higgs mass of 500 GeV,   
the light $B_H$ region ($M_B \sim 180$ GeV) has less than 20\% 
contribution from annihilation to $t\bar{t}$, with the remainder shared between
$W^+ W^-$ and $ZZ$ in accordance with the equivalence theorem; whereas for 
a heavy $B_H$ ($M_B\sim$ 350 GeV) there is a negligible contribution from 
$t\bar{t}$.  These relative contributions from different final states can 
be read directly from Fig.~\ref{fig:xsection}.

We expect effects of 
coannihilation with other level-1 states to be small
due to larger mass splittings between the modes as compared with 
those in 5D~\cite{Servant:2002aq,Kong:2005hn}, and we do not include these in our analysis.
Our results are relatively insensitive to exotic Higgs decays since their
contributions to the total width of the Higgs are small.  Moreover, they 
are also mostly independent of the rest of the KK spectrum of the
6DSM.  Recall that only annihilation to top quarks involves any additional 
heavy modes, and that this contribution is subdominant over most of the 
parameter space. 

\section{Astrophysical Detection} \setcounter{equation}{0}

Efforts to detect dark matter particles with astrophysical experiments are 
often classified as direct or indirect detection. Direct detection experiments are 
those which attempt to observe particles scattering elastically with the detector, 
whereas indirect detection efforts attempt to observe the dark matter annihilation products~\cite{dmreview}. 

\subsection{Direct Detection}

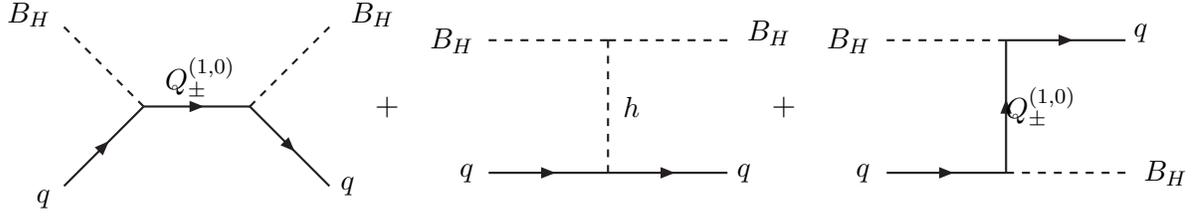
\begin{figure*}[t]
\unitlength=1.0 pt
\SetScale{1.0}
\SetWidth{0.8}      
\begin{center}
\begin{picture}(400,80)(10,20)
\DashLine(20,80)( 50,50){3}
\DashLine(90,50)( 120,80){3}
\ArrowLine(20,20)(50,50)
\ArrowLine(90,50)(120,20)
\ArrowLine(50,50)(90,50)
\Text(147,50)[r]{{\large +}}
\DashLine(180,75)(225,75){3}
\DashLine(225,75)(270,75){3}
\DashLine(225,25)(225,75){3}
\ArrowLine(180,25)(225,25)
\ArrowLine(225,25)(270,25)
\Text(297,50)[r]{{\large +}}
\DashLine(330,75)(375,75){3}
\DashLine(420,25)(375,25){3}
\ArrowLine(330,25)(375,25)
\ArrowLine(375,25)(375,75)
\ArrowLine(375,75)(420,75)
\Text(15,15)[r]{$q$}
\Text(15,85)[r]{$B_H$}
\Text(175,25)[r]{$q$}
\Text(175,75)[r]{$B_H$}
\Text(325,25)[r]{$q$}
\Text(325,75)[r]{$B_H$}
\Text(85,61)[r]{$Q_\pm^\one$}
\Text(130,20)[r]{$q$}
\Text(145,85)[r]{$B_H$}
\Text(280,25)[r]{$q$}
\Text(295,78)[r]{$B_H$}
\Text(445,25)[r]{$B_H$}
\Text(430,78)[r]{$q$}
\Text(238,50)[r]{$h$}
\Text(403,50)[r]{$Q_\pm^\one$}
\end{picture}
\end{center}
%
\caption{Tree-level diagrams for the elastic scattering of the $B_H$ with quarks.}
\label{fig:direct}
\end{figure*}

In this section, we discuss
the prospect for the direct detection of spinless photon dark matter.
A spinless photon can scatter elastically with a quark through the exchange 
of a KK-quark or a Higgs boson (see Fig.~\ref{fig:direct}). 
The leading term in the amplitude due to Higgs exchange is given in the non-relativistic limit by
\begin{equation}
{\cal M}_h = i \, \frac{g_Y^2}{2}\frac{m_q}{m_h^2} \, \bar{q}q  \, ,
\label{gammaq}
\end{equation}
where $q$ is a quark field of mass $m_q$.
Similarly the amplitude for KK quark exchange is given by
\begin{eqnarray}\label{equ:elasticquark}
{\cal M}_Q &=& - i \frac{g_Y^2}{4} \Big [
( y_L^2 + y_R^2  ) \Big ( \frac{m_q -M_B}{(m_q-M_B)^2 - M_Q^2} +  
                          \frac{m_q +M_B}{(m_q+M_B)^2 - M_Q^2}  \Big )  \bar{q} \, \gamma^0 \, q \nonumber \\
&&+ 2 M_Q y_L y_R \sin 2 \alpha \Big ( \frac{1}{(m_q-M_B)^2 - M_Q^2}
                                   + \frac{1}{(m_q+M_B)^2 - M_Q^2}  \Big )  \bar{q}q
 \Big ] \, .
\end{eqnarray}
Summing ${\cal M}_h$ and ${\cal M}_Q$, we obtain
\begin{equation}
\langle {\cal M} \rangle = {\cal C}_q \langle \bar{q}q\rangle  \, ,
\end{equation}
where we have combined terms using $\bar{q}\gamma^0 q \approx \bar{q}q$ and $\bar{q}\gamma^5 q \approx 0$ 
which hold in the non-relativistic limit. $\langle \,\,\, \rangle$ denotes an average and sum over the spins of the initial and final state quarks respectively.
The coefficient ${\cal C}_q$ may be read directly from the matrix elements~(\ref{gammaq})
and~(\ref{equ:elasticquark}),
\begin{eqnarray}\label{eqn:c}
{\cal C}_q &=& 
\frac{g_Y^2}{4} \left[
m_q ( y_L + y_R )^2 \left( \frac{ 1}{M_Q^2 - (m_q-M_B)^2 } +  
                           \frac{1}{M_Q^2 - (m_q+M_B)^2 }  \right) \right.
\nonumber \\ [0.3em]
&+& \left. M_B ( y_L^2 + y_R^2 )  \left(  \frac{1}{M_Q^2 - (m_q+M_B)^2 }
                                 - \frac{1}{M_Q^2 - (m_q-M_B)^2 }  \right)+ \frac{2 m_q}{m_h^2}  \right]\;.
\end{eqnarray}
The propagators in this expression can be expanded to linear order in $m_q$
to obtain
\begin{equation}
{\cal C}_q\approx \frac{g_Y^2}{2} m_q \left(\frac{1}{m_h^2} 
+ \frac{(y_L+y_R)^2}{M_Q^2-M_B^2} + \frac{2\left(y_L^2+y_R^2\right) M_B^2}
{\left(M_Q^2-M_B^2\right)^2}\right) ~.
\end{equation}
The expression in Eq.~(\ref{eqn:c}) diverges for $M_Q = | M_B \pm m_q |$.
Given that 
we are ultimately interested in elastic scattering off nucleons, the top 
quark contributes only at one loop through the effective coupling of a pair 
of $B_H$s to two gluons. 
For simplicity, 
we treat the contribution from the top quark in the same way as that from the $b$ or $c$ quarks.
The validity of this procedure would need to be checked 
by a full loop calculation of $B_H$-gluon elastic scattering, which would allow one to assess
whether there are any resonance effects.

Note that effects from electroweak mass mixing that flip the chirality of the (1,0) quarks
(proportional to $y_L y_R$ in the above equation)
are of the same order as the pieces that flip the chirality of the external quark lines,
and may not be neglected.\footnote{In the 5D case, we expect that similar terms, which have been
omitted so far in the literature, 
will increase the contribution from KK quark exchange to 
the elastic scattering cross section.}
As anticipated from the discussion in Sec.~\ref{sec:xsec} of higher-dimension operators 
contributing to the elastic scattering process, the entire $B_H$-quark elastic scattering cross section is
proportional to $m_q$.   

The matrix element $\langle \bar{q}q\rangle$ of quarks in a nucleon state can be evaluated~\cite{Jungman:1995df} to
obtain
\begin{equation}
\langle \bar{q}q\rangle = \frac{m_{p,n}}{m_q} f^{p,n}_{T_q}~~{\rm (light ~quarks)}~;~~
\langle \bar{q}q\rangle = \frac{2}{27}\frac{m_{p,n}}{m_q} f^{p,n}_{TG}~~{\rm (heavy ~quarks)}~.
\end{equation}
Summing over quark flavors, we arrive at the $B_H$-nucleon couplings: 
\begin{equation}
f^{B_H}_{p,n} = m_{p,n} \sum_{q=u,d,s} \frac{{\cal C}_q }{m_q} f^{p,n}_{T_q} 
    + \frac{2 m_{p,n}}{27} f^{p,n}_{TG} \sum_{q=c,b,t} \frac{ {\cal C}_q}{m_q} \, ,
\end{equation}
where the quantities $f^{p,n}_{T_q}$ have been measured to be 
$f^p_{T_u}=0.020\pm 0.004$,  $f^p_{T_d}=0.026\pm 0.005$,  
$f^p_{T_s}=0.118\pm 0.062$, $f^n_{T_u}=0.014\pm 0.003$,  $f^n_{T_d}=0.036\pm 0.008$ 
and  $f^n_{T_s}=0.118\pm 0.062$ \cite{nuclear}. 
The first term in this expression corresponds to interactions with quarks 
in the target nucleon, whereas the second term results from interactions 
with gluons through a quark or heavy quark loop. $f^p_{TG}$ is given by 
$1-f^p_{T_u}-f^p_{T_d}-f^p_{T_s} \approx 0.84$ and analogously, 
$f^n_{TG} \approx 0.83$.

\begin{figure}[t]
\centerline{
\includegraphics[width=.8 \textwidth]{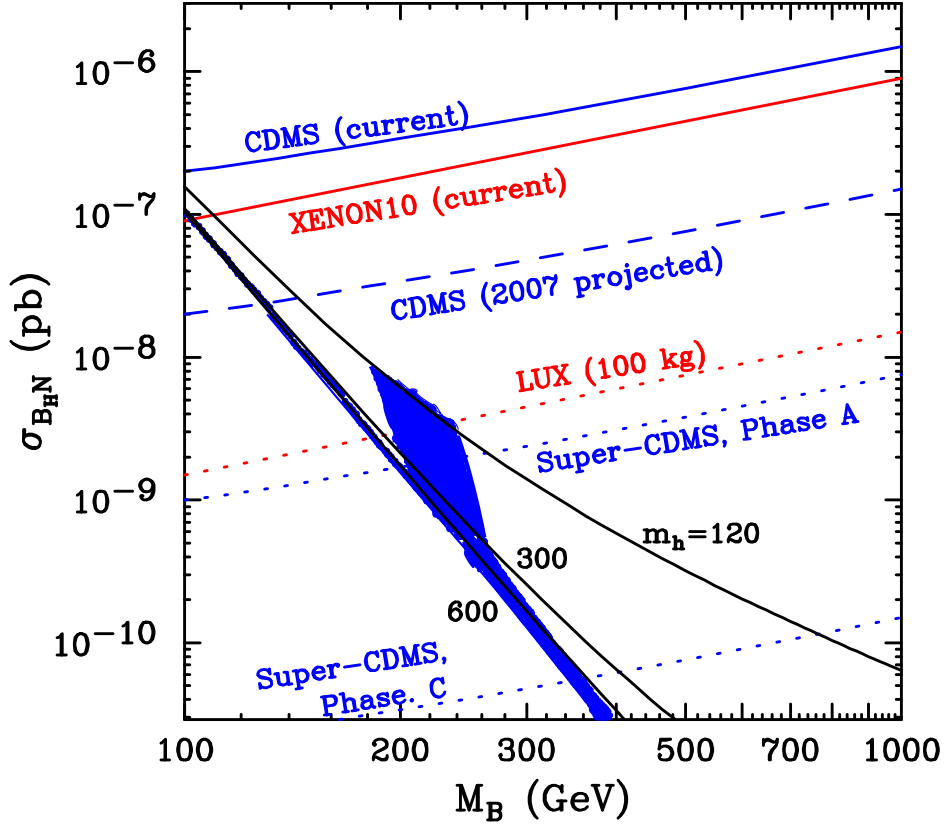} 
}
\caption{Prospects for the direct detection of $B_H$ dark matter. 
Predicted cross sections are shown as black solid lines for Higgs masses of 120, 300 and 600 GeV. 
Current constraints are shown as blue (CDMS) and red (XENON) solid lines. 
The dashed blue line denotes the near term projection from the CDMS experiment. The dotted lines represent longer term projections. Shown as a filled blue region is the parameter range in which the observed abundance of dark matter can be generated in this model.}
\label{fig:xsec_direct}
\end{figure}
The total $B_H$-nucleus cross section at zero momentum transfer is given by
\begin{equation}
\sigma = \frac{m^2_N}{4 \pi (M_B + m_N)^2} \bigg(Z f^{B_H}_{p} + (A-Z) f^{B_H}_{n}\bigg)^2 \, ,
\end{equation}
where $m_N$, $Z$ and $A$ are the mass, atomic number and atomic mass of the target nuclei. 
Although the experimental sensitivities and limits are often described in terms of the 
dark matter elastic scattering with nucleons, one should keep in mind that the nuclear form factors
may need to be taken into account.

Note that there is no spin-dependent contribution to the elastic scattering cross section. 
This is in contrast with the 5D case, where the spin-dependent $B^{(1)}_\mu$-nucleus 
elastic scattering cross section is 
typically three or four orders of magnitude larger than the corresponding spin-independent 
cross section~\cite{Cheng:2002ej,Servant:2002hb}, and only the average over the nucleons
inside the nucleus suppresses the spin-dependent effects.

In Fig.~\ref{fig:xsec_direct} we compare the spin-independent elastic scattering cross section 
of the $B_H$ to the current and projected sensitivities of direct detection experiments. 
At present, the strongest limits have been placed 
by the XENON~\cite{Angle:2007uj} and CDMS~\cite{cdms} collaborations. 
These constraints are, however, not yet sensitive to the range of cross sections predicted 
in this model. Only with future experimental programs, such as the first phase of Super-CDMS 
or a 100 kilogram version of LUX, will direct detection experiments begin to reach 
the sensitivity needed to test this model. To test the region with $M_B$ of order
several hundred GeV and larger, the full phase-C of super-CDMS or a multi-ton 
liquid noble detector will likely be required~\cite{future}. 

\subsection{Indirect detection} \setcounter{equation}{0}

Efforts to detect the annihilation products of dark matter particles 
in the form of gamma rays, antimatter and neutrinos are collectively 
known as indirect detection. 
In this section, we discuss the prospects for the indirect detection 
of spinless photon dark matter.

Dark matter particles annihilating in the galactic halo or in 
dark matter substructures may potentially 
generate observable fluxes of annihilation products in the form of gamma rays, positrons, 
anti-protons or anti-deuterons. The prospects for searches of such particles depend strongly 
on unknown astrophysical inputs, such as the distribution of dark matter and 
the structure of galactic magnetic fields. The only particle physics inputs 
which are relevant to gamma ray and antimatter searches for dark matter are 
particle's mass, annihilation cross section in the low velocity limit and 
the species of Standard Model particles that are generated in those annihilations.

The low velocity cross section for spinless photon annihilations is dictated 
by the relic abundance calculation to be $\sigma v \approx 3 \times 10^{-26}$ cm$^3$/s $\approx$1 pb. These annihilations largely result 
in the production of gauge and Higgs boson pairs. This is very similar to the characteristics found 
for a wino-like or higgsino-like neutralino, leading to very similar prospects and 
signatures in gamma ray and antimatter based dark matter searches. 
Instead of repeating the phenomenology of these indirect detection channels here, 
we refer the reader to previous studies on the subjects of dark matter searches 
with gamma rays~\cite{gammarays} and antimatter~\cite{antimatter}.
We will, however, mention briefly the differences found between the 6D and 5D cases 
regarding these. 

In the case of 5D, 
the $B^{(1)}_{\mu}$ annihilations generate mostly charged lepton pairs 
(approximately 20\% to each family). In addition to the standard gamma ray spectrum 
from cascade decays and fragmentation, the electron-positron pairs produce a harder 
gamma ray spectrum via final state radiation. The tau pairs produced also generate 
a harder spectrum through their decays~\cite{Bergstrom:2004cy}. 
In 6D, the annihilations to gauge and Higgs bosons do not result in such a hard spectrum.

In addition, annihilations to electron-positron pairs in and 
other charged leptons in 5D result in a particularly hard spectrum of positrons 
in the cosmic ray spectrum \cite{Cheng:2002ej,kribspos}. 
The contribution from annihilations to $W^+ W^-$ in the 6DSM is also somewhat hard, 
but much less so than is found in the 5D case. 
As with the gamma ray spectrum, the positron spectrum resulting from dark matter annihilations 
in this model more closely resembles the signal predicted from neutralino annihilations than 
from the case of 5D models.

Dark matter particles which undergo elastic scattering with nuclei in the Sun or Earth can become gravitationally 
bound to these bodies, and accumulate in their cores. Once captured in sufficient numbers, 
they can annihilate efficiently, producing a sizable flux of energetic Standard Model particles. 
Of these annihilation products, only neutrinos can escape from the Sun or Earth and potentially be observed~\cite{neutrino}.

The capture rate of dark matter particles depends on their elastic scattering cross section with nuclei. 
Unfortunately, this cross section is rather small in the model considered here. 
Over the entire range of parameters considered here, the elastic scattering cross section is never larger than
$\sim 10^{-7}$ pb, which leads to less than one neutrino being 
observed from dark matter annihilations in the Sun per ten years in a kilometer-scale 
experiment~\cite{halzen}. The rate from the Earth is even smaller.
This is very different from the neutrino rate predicted in the 5D case. 
The reason for this distinction is that spin-dependent scattering is significant in 5D, 
leading to typical rates of $\sim 0.1-100$ per square kilometer per year~\cite{Hooper:2002gs}. 

\section{Conclusions} \setcounter{equation}{0}

Despite the experimental successes of the Standard Model, it does not contain a viable candidate for dark matter. This absence is one of the strongest motivations for the existence of physics beyond the Standard Model.
In particular, dark matter is a primary motivation for supersymmetry since models with 
R-parity conservation can provide a viable dark matter candidate. 
Recently, there has been greater attention placed on other types of dark matter candidates, including those found in models with 
universal extra dimensions, where the stability of dark matter is due to a discrete
symmetry called KK parity. 
In the minimal model with one universal extra dimension, 
dark matter typically consists of a KK excitation of the hypercharge gauge boson.
This 5D KK dark matter has strikingly different phenomenology from neutralinos in supersymmetric models~\cite{gabi}.
Models with one universal extra dimension may also contain other viable dark matter candidates, including
the KK modes of the graviton \cite{Feng:2003xh} and right-handed neutrinos~\cite{Matsumoto:2006bf}.
Certain models with two universal extra dimensions, where the dark matter particle is a KK mode of 
the hypercharge vector boson or right-handed neutrinos, have also been investigated \cite{Servant:2002aq,Mohapatra:2002ug}. 

In this paper we have studied the possibility of KK dark matter in the 6DSM \cite{Ponton:2005kx}, which is 
the minimal model with two universal extra dimensions.
The lightest KK-parity odd state is a spin-0 excitation of the hypercharge boson, $B_H$, referred to as 
the spinless photon. We have computed annihilation cross sections necessary for the calculation of relic density in this model, and found the regions of parameter space in which the measured abundance of dark matter is generated.

Unlike KK dark matter in the 5D case, $B_H$ annihilations into fermion final states is 
helicity suppressed because $B_H$ has spin 0.
Thus, all fermion final states other than top quarks are negligible, and 
final states with bosons are dominant. In order for the $B_H$ to sufficiently annihilate and 
to generate the desired thermal relic density its mass must satisfy $M_B\lesssim$ 500 GeV.  
In the 6DSM this corresponds to a
compactification scale of $1/R\lesssim$ 600 
GeV, which is considerably smaller than the range favored in the 5D case.

It is tempting to compare this upper limit with the lower limit from
searches at the Tevatron, of almost 300 GeV \cite{Dobrescu:2007xf}. However,
one should
keep in mind that perturbations of the mass spectrum due to localized operators
could change the limit from relic abundance independently of the collider
limits,
as they depend on different (1,0) masses. The limits from electroweak
observables
have not been computed in the 6DSM, and are likely in any case to be sensitive
to
contributions from the unknown physics at the cutoff scale.

We have also studied the prospects for observing $B_H$ dark matter in direct and 
indirect dark matter experiments. 
We find that the elastic scattering cross section of this particle with nuclei 
is completely spin-independent, 
and is smaller than the current sensitivity of direct detection experiments. 
Only the next-generation experiments will start probing significant
regions of the parameter space.
Moreover this small spin-independent cross section results in a prediction of very small
rates at neutrino telescopes.  
The phenomenology of the spinless photon in the context 
of astrophysical detection resembles neutralino dark matter in many respects, and 
is distinctively different from KK dark matter in models with one universal extra 
dimension.\footnote{The collider phenomenology of the 6DSM, on the other hand, is quite different
from either supersymmetry or one universal extra dimension. It includes distinctive 
multi-lepton plus photon signatures \cite{Dobrescu:2007xf}, and 
multiple $t\bar{t}$ resonances \cite{Burdman:2006gy}.}

These conclusions could potentially be modified once other effects
are considered. In particular, a pair of $B_H$'s may annihilate via an $s$-channel
(2,0) Higgs exchange, and if the masses are near the resonance 
that could be a large effect even though 
the coupling of the (2,0) modes to standard model particles are suppressed (a similar situation
occurs in the case of one universal extra dimension \cite{Kakizaki:2005en}).
Furthermore, the mixing of $B_H$ 
with the spinless $Z$-boson, the next lightest KK mode, may be
an important effect for small $M_B$.  In this scenario we expect to 
see a non-negligible increase of the $B_H$ mass in order for the 
relic abundance to remain consistent with WMAP measurements.  
Coannihilations tend to increase the range of dark matter masses which yield
the measured abundance of dark matter regardless of the Higgs mass~\cite{Kong:2005hn}.
Higher-order corrections to the annihilation cross section are known to be non-negligible, 
especially in the case of coannihilation with colored particles and when quarks are in 
the final state. In our study, however, this only applies to 
annihilation to $t\bar{t}$, which is somewhat suppressed compared to annihilation 
to $W^+W^-$, $ZZ$ and $hh$. 

Dark matter candidates other than the spinless photon may be possible in the 6DSM  
if the (1,0) spectrum is modified by localized operators. These include the (1,0) modes of the 
graviton, right-handed neutrino, Higgs boson, or electroweak bosons.
We leave a further exploration of these possibilities to future studies.

\bigskip

{\bf Acknowledgments:} \ Many thanks to Jonathan Feng, Geraldine Servant and especially 
to Tim Tait for useful comments.
Fermilab is operated by Fermi Research Alliance, LLC under Contract No. 
DE-AC02-07CH11359 with the United States Department of Energy. 
DH is also supported by NASA grant NAG5-10842.

 \vfil \end{document}